\documentclass[aps,prd,preprint,groupedaddress]{revtex4-2}
\usepackage{amsmath}
\usepackage{graphicx}
\usepackage{todonotes}

\begin{document}
\begin{titlepage}
\title{A simple model to include initial-state and hot-medium effects in the computation of quarkonium nuclear modification factor} 
\author{Miguel \'{A}ngel \surname{Escobedo}}
\author{Elena G. Ferreiro}
\affiliation{Instituto Galego de F\'{i}sica de Altas Enerx\'{i}as (IGFAE), Universidade
de Santiago de Compostela, E-15782, Galicia, Spain.}
\date{\today}
\begin{abstract}
Quarkonium suppression is one of the more useful observables to obtain information about the hot medium created in ultrarelativistic heavy-ion collisions. 
In this manuscript, we discuss a simple way to implement both the initial-state effects and the hot-medium evolution, and to compute the quarkonium nuclear modification factor if the survival probability for a bound state at a given energy density is known.
Based on the Glauber model, the temperature of the evolving medium and the centrality dependence of the nuclear modification factor will be analysed.
\end{abstract}
\maketitle
\end{titlepage}
\section{Introduction}
Quarkonium suppression was proposed in \cite{Matsui:1986dk} as a signal of a formation of a quark-gluon plasma. Over the years, the models for the description of quarkonium evolution in a hot medium have been refined \cite{Laine:2006ns,Brambilla:2008cx,Escobedo:2008sy,Beraudo:2007ky,Margotta:2011ta,Strickland:2011mw,Akamatsu:2014qsa,Katz:2015qja,Blaizot:2017ypk,Du:2017qkv,Yao:2017fuc,Yao:2018nmy,Yao:2018sgn,Yao:2020xzw,Yao:2020eqy} and our theoretical description is now very different to the original idea. However, to obtain a realistic description of quarkonium in a heavy-ion collision, it is not enough to understand how quarkonium interacts with a medium. In the following, we mention just some of the relevant issues:
\begin{itemize}
\item We need to take into account the time evolution of the medium itself. Relativistic hydrodynamics is one of the most common frameworks \cite{Heinz:2013th} as it gives an excellent description of the observables related to the bulk of the medium. If we assume that the nucleus properties are homogeneous in the transverse direction, we obtain the Bjorken evolution \cite{Bjorken:1982qr}. In any case, we need to add as input some information about the initial state of the medium, usually the initial energy density at any point in the transverse plane, and the initial time at which a hydrodynamic description of the medium is valid.
\item Naively, we would expect that we could predict what would happen to quarkonium in a heavy-ion collision if the medium would not modify quarkonium’s population from extrapolating proton-proton data. However, this is very far from reality. It is quite non-trivial even to predict proton-nucleus collisions, in particular due to the presence of initial state effects that lead to the modification of the nuclear Parton Distribution Functions (nPDFs). Several models exist for this task \cite{Capella:2011vi,Eskola:2016oht,Kusina:2017gkz,Ma:2015sia}.

Here we propose to use the model of \cite{Capella:2011vi}, based on Pomeron interactions. In this model, 
the origin of the modification of the nuclear ratios is related to multiple scattering with nucleons. 

This results in a reduction of the corresponding cross sections, commonly called shadowing. 
This simple model offers the advantage of being completely analytical and provides a reasonable description of proton-nucleus data.

\end{itemize}

Our aim in this manuscript is to provide a simple method to compute the nuclear modification factor in heavy-ion collisions, taking into account initial nuclear matter and hot medium effects. 
To do so, we will assume that the following information is available:
\begin{itemize}
\item The initial state of a heavy quark pair after a proton-proton collision.
\item The probability that this heavy quark pair forms a given bound state provided that it interacts with an evolving medium which initially has an energy density $\epsilon_i$.
\end{itemize}

First, we will compute how the initial temperature depends on the position on the transverse plane. This temperature depends on the medium's energy density, that we assume proportional to the particle multiplicity, {\it i.e.} to the number of pions. We describe the multiplicity using the model of \cite{Capella:2011vi} which takes into account nuclear shadowing effects. We argue that to a very good approximation, this is equivalent to assuming that the initial energy density at a given point in the transverse plane is proportional to the density of participant nucleons at that point. 
This statement is true at least at nowadays collinear energies. To illustrate this, we will show our results for the temperature behaviour in 3 different cases, {\it i.e.} considering it to be proportional to the number of participants, to the number of collisions and to the number of pions, once shadowing effects are taken into account. 
Then, we will illustrate how to compute the quarkonium's nuclear modification factor in heavy-ion collisions including initial and hot nuclear matter effects. We remark that our aim is not to model quarkonium interaction with a medium but to provide a prescription to include initial effects and temperature evolution for a quarkonium computation in a hot environment. This prescription has the advantage that the effects of shadowing can be incorporated in a straight-forward way using analytical formulas.

The outline of the manuscript is as follows. In section \ref{sec:Glauber}, we review some aspects of the Glauber model useful to set the scope and notation of our computation. In section \ref{sec:model}, we review the initial state model of \cite{Capella:2011vi}. In section \ref{sec:temp}, we study the initial energy density and the temperature of the medium. 
In section \ref{sec:ahq}, we show how to implement our model for the computation of the nuclear modification factor.
In section \ref{sec:raa}, we illustrate our prescription with some examples in which we include both hot and initial nuclear matter effects on quarkonium. Finally, in section V, we give our conclusions.  

\section{Glauber model review}
\label{sec:Glauber}
For completeness, we review in this section the Glauber model \cite{Glauber:1970jm} in the optical approximation \cite{Bialas:1977pd}. This is needed in order to fix the notation. Many reviews of the Glauber model can be found in the literature, in our case we follow closely the description given in \cite{Miller:2007ri}.

The optical Glauber model is used to extrapolate the known properties of the collisions between nucleons to describe collisions of large nuclei made of many nucleons. For all practical purposes within this manuscript, we take the following physical assumptions:
\begin{itemize}
\item The eikonal approximation is used for the nucleons inside a nucleus. This means that for the purposes of studying the nucleus-nucleus collision we can consider that the nucleons do not have transverse motion. This approximation is correct up to subleading corrections of order $\frac{\Lambda_{QCD}}{\sqrt{s}}$.
\item The interaction between nucleons is local, therefore, two nucleons of different nucleus can only interact if they are at the same point in the transverse plane during the collision.
\item Collisions between nucleons are independent processes. In other words, the probability of collision between two nucleons does not depend on whether the rest of the nucleons have collided or not. 
\end{itemize}
The function that contains all the required information of the nucleus is the density of nucleons inside a nucleus, which is assumed to be proportional to the nuclear charge density. We focus on the case of a spherically symmetric nucleus and we use a Woods-Saxon profile
\begin{equation}
\rho_A(r)=\frac{\rho_0}{1+e^{\frac{r-c}{\xi}}}\,,
\end{equation}
where for the case of lead $c=6.62\,\rm{fm}$ and $\xi=0.546\,\rm{fm}$ (These values are taken from \cite{DeJager:1987qc}). $A$ is the nucleon number and $\rho_0$ is chosen such that
\begin{equation}
\int_0^\infty\,d^3r\rho_A(r)=A\,.
\end{equation}
An analytical expression can be found for $\rho_0$
\begin{equation}
\rho_0=\frac{-A}{8\pi\xi^3Li_3(-e^{\frac{c}{\xi}})}\,,
\end{equation}
where $Li$ is a polylogarithm. From the density of nucleons we can compute the thickness function, which informs of the density of nucleons at a given position in the transverse plane for any value of the longitudinal coordinate
\begin{equation}
T_A(x,y)=\int_{-\infty}^{\infty}\,dz\rho_A(x,y,z)\,.
\end{equation}
Since we are interested in the collision of a nucleus with nucleon number $A$ with another with nucleon number $B$, it is useful to define the overlap function, {\it i.e.} the density of pairs of nucleons from different nuclei but at the same point in the transverse plane -so that they might interact. This can be defined for a given value of the impact parameter $b$ as
\begin{eqnarray}
T_{AB}(b) & = &\int\,d^2s T_A\left(\bf{s}+\frac{\bf{b}}{2}\right)T_B\left(\bf{s}-\frac{\bf{b}}{2}\right)\nonumber \\
& = & \int\,d^2s\,dz_1\,dz_2\rho_A\left(\sqrt{\left(\mathbf{s}+\frac{\mathbf{b}}{2}\right)^2+z_1^2}\right)\rho_B\left(\sqrt{\left(\mathbf{s}-\frac{\mathbf{b}}{2}\right)^2+z_2^2}\right)\,.
\end{eqnarray} 
Note that in the previous equation we use a coordinate system for the transverse plane such that $\mathbf{s}=\mathbf{0}$ corresponds to the center of the colliding region when the two nuclei are identical. This is not the standard choice in the literature, where it is common to choose an origin of coordinates which coincides with the center of one of the nuclei. Our motivation is that we expect many of the properties of the collision to exhibit a central plateau and we want a coordinate system whose origin is closed to the center of the plateau.

Until now we have only described the geometry of the collision. Another required input is the cross section for inelastic collision between two nucleons, which in order to simplify the notation we represent with $\sigma$. At the LHC, the value $\sigma=64\,\rm{mb}$ is taken for $\sqrt{s}=2.76\,\rm{TeV}$ collisions while the value $\sigma=70\,\rm{mb}$ is taken for $\sqrt{s}=5.02\,\rm{TeV}$ collisions. With this we can compute the probability of having $n$ nucleon collisions. Since we assume that the collisions are independent processes it must follow a binomial distribution. We consider the number of $n$ pairs that can be made out of $A$ nucleons from one nucleus and $B$ from the other, then we consider the probability that these $n$ pairs collide while the rest of possible pairs do not. This reasoning leads to the following expression
\begin{equation}
P(n,b)=\left(\begin{array}{c}
AB \\
n \\
\end{array}\right)\left(\frac{T_{AB}(b)\sigma}{AB}\right)^n\left(1-\frac{T_{AB}(b)\sigma}{AB}\right)^{AB-n}\,.
\end{equation}
Then the probability that there is at least an inelastic collision is
\begin{equation}
P_{inel}(b)=\sum_{n=1}^\infty P(n,b)=1-P(0,b)=1-\left(1-\frac{T_{AB}(b)\sigma}{AB}\right)^{AB}\,,
\label{eq:Pinel}
\end{equation}
and the mean value of the number of collisions is
\begin{equation}
N_{coll}=\sum_{n=1}^\infty nP(n,b)=T_{AB}(b)\sigma\,.
\end{equation}
Another quantity which is used to characterise heavy-ion collisions is the number of participants, {\it i.e.} the total number of nucleons that collide at least once. To compute it we can use the fact that, in the optical Glauber model, collisions are assumed to be independent. Thus we can compute the number of participants in the following way:
\begin{itemize}
\item Take one nucleon from the nucleus with mass number $A$.
\item Compute the probability that this nucleon has an inelastic collision with the other nucleus. To do so we can use eq.~(\ref{eq:Pinel}) adapted to the $pB$ case ($P(\mathbf{s})_{inel,pB}=1-\left(1-\frac{T_{B}\left(\mathbf{s}-\frac{\mathbf{b}}{2}\right)\sigma}{B}\right)^{B}$).
\item Then we sum over the distribution of nucleons in the nucleus with mass number $A$.
\item We repeat the process but now studying the probability that a nucleon of nucleus $B$ interacts with nucleus $A$. 
\end{itemize}
Therefore, we arrive at the following expression 
\begin{equation}
\begin{split}
N_{part}=\int\,d^2sT_A\left(\mathbf{s}+\frac{\mathbf{b}}{2}\right)\left(1-\left(1-\frac{T_{B}\left(\mathbf{s}-\frac{\mathbf{b}}{2}\right)\sigma}{B}\right)^{B}\right)+\\
+\int\,d^2sT_B\left(\mathbf{s}-\frac{\mathbf{b}}{2}\right)\left(1-\left(1-\frac{T_{A}\left(\mathbf{s}+\frac{\mathbf{b}}{2}\right)\sigma}{A}\right)^{A}\right)\,.
\end{split}
\end{equation}

In many cases, experimental results are given in terms of centrality classes. It is assumed that the multiplicity increases as the impact parameter decreases. Then, for example, the $10\,\%$ of collisions with higher multiplicity corresponds to the $10\,\%$ more central collisions. Therefore, the $0-10$ centrality window includes collisions from $b_0=0$ up to another impact parameter that we call $b_{10}$. The numerical value of $b_{10}$ can be computed in the following way. Let us assume that the multiplicity is proportional to the inelastic cross section, given in eq.~(\ref{eq:Pinel}). It is convenient to simplify eq.~(\ref{eq:Pinel}) using that $AB\gg 1$
\begin{equation}
P_{inel}(b)\sim 1-\mathrm{e}^{-T_{AB}(b)\sigma}\,.
\end{equation}
It follows that the total inelastic cross section is
\begin{equation}
\sigma_{inel}=\int\,d^2b\left(1-\mathrm{e}^{-T_{AB}(b)\sigma}\right)\,.
\end{equation}
Then $b_{10}$ can be determined solving the following equation
\begin{equation}
\frac{2\pi\int_0^{b_{10}}\,dbb\left(1-\mathrm{e}^{-T_{AB}(b)\sigma}\right)}{\sigma_{inel}}=0.1\,.
\end{equation}
Analogously, the centrality window $x-y$ includes collisions in the range of impact parameters between $b_x$ and $b_y$ such that
\begin{equation}
\frac{2\pi\int_{b_x}^{b_y}\,dbb\left(1-\mathrm{e}^{-T_{AB}(b)\sigma}\right)}{\sigma_{inel}}=\frac{y-x}{10}\,.
\end{equation}

\section{Initial-effect model} 
\label{sec:model}

As it is commonly assumed, the quarkonium production, primarily formed by gluon-gluon fusion, will suffer the shadowing effects that modified the parton distribution functions in a nucleus and affect global multiplicities.
Several models, mostly based on data parametrization at a given $Q^2$ followed by Dokshitzer–Gribov–Lipatov–Altarelli–Parisi (DGLAP) evolution, exist \cite{Eskola:2016oht,Kusina:2017gkz}. 
For our purpose, it is more useful to implement  the analytical shadowing model based on Glauber-Gribov theory \cite{Capella:2011vi}.
 In this model, the interaction of two high-energy colliding objects is described by Pomeron exchanges. The Pomeron is a Regge pole accompanied by the cuts associated to exchanges in the $t$-channel. 
 It is represented by a ladder-type diagram. Cutting the ladder leads to the production of secondary particles, which correspond to inelastic interactions. 
 Shadowing corrections appear through multiple scattering contributions.
In parton language, the Pomeron exchange is equivalent to the exchange of a pair of gluons. Partons of a fast nucleus with small relative momenta overlap in the longitudinal space and can interact. For example, two ladders of partons from different nucleons can fuse into one ladder. This corresponds to the diagram with triple-Pomeron interaction. From the point of view of Gribov, 
 diagrams involving triple-Pomeron interactions are related to large-mass intermediate states 
 so the corresponding couplings  
 can be determined from experimental data on inclusive diffractive processes. Therefore, in Gribov theory the diffractive production in deep inelastic scattering processes on nucleus and the nuclear shadowing effects have the same origin:
the diffractive cuts give a positive contribution to diffraction and a negative one -the shadowing corrections- to the total cross section through the
Abramovskii-Gribov-Kancheli (AGK) cutting rules \cite{Abramovsky:1973fm}.

In order to relate diffraction on nucleons with nuclear shadowing, the $\gamma^*{\rm nucleus}$ cross section can be expanded in a multiple scattering series containing the contribution from 1,2,... scatterings, $\sigma_{\gamma^*A}=\sigma_{\gamma^*A}^{(1)}+\sigma_{\gamma^*A}^{(2)}+...$

$\sigma_{\gamma^*A}^{(1)}$ is just equal to $A \sigma_{\gamma^*\rm nucleon}$, the Glauber elastic contribution. The amount of the shadowing corrections corresponding to two-scattering contribution obeys
\begin{equation}
\label{eq14}
    f(x,Q^2)=4\pi
\int _{M^2_{min}}^{M^2_{max}}dM^2 \left. \frac{1}{\sigma_{\gamma^*{\rm nucleon}}}
\frac{d\sigma^{\mathcal{D}}
_{\gamma^*{\rm nucleon}}}{dM^2dt}\right\vert_{t=0}\, .
\end{equation}
Eq.~(\ref{eq14}) represents the first contribution to nuclear
shadowing originated by double scattering with two target nucleons. Convoluted with the normalised nuclear profile function, it gives a negative contribution to the total cross section, 
$\sigma_{\gamma^*A}^{(2)}=-A(A-1)\int d^2b\, T^2_A(b)\, f(x,Q^2)$.

The ratio of the differential cross section for diffractive dissociation of the virtual photon over the total $\gamma^*{\rm nucleon}$ cross section
$\frac{1}{\sigma_{\gamma^*{\rm nucleon}}}
\frac{d\sigma^{\mathcal{D}}_{\gamma^*{\rm nucleon}}}{dM^2dt}\vert_{t=0}$ that appears in the above equation is directly related to
the ratio of the triple-Pomeron cross section over the single Pomeron one
$\frac{1}{\sigma_P}\left.\frac{d\sigma^{PPP}}
{dydt}\right\vert_{t=0}$ and can be interpreted as the reduction due to the interaction among the partons. 

Thus, in terms of triple-Pomeron contributions, eq.~(\ref{eq14}) reads
\begin{equation}
F(y,p_T)
=4\pi\int_{y_{min}}^{y_{max}}dy\ \frac{1}{\sigma_P}\left.\frac{d\sigma^{PPP}}
{dydt}\right\vert_{t=0}
=\int_{y_{min}}^{y_{max}}dy\ C \exp (\Delta \times y)
\, ,
\end{equation}
where we have introduced
the standard triple-Pomeron formula which involves 
the Pomeron intercept and its residue, $4\pi\frac{1}{\sigma_P}\left.\frac{d\sigma^{PPP}}
{dydt}\right\vert_{t=0}=C \exp (\Delta \times y)$, with
$C=\frac{g_{pp}^P r_{PPP}}{4}$, being $g_{pp}^P$ the Pomeron-proton coupling and $r_{PPP}$ the triple-Pomeron coupling, both evaluated at $t=0$, $C = 0.023$~fm$^2$. The value of  $\Delta$ is related to the value of the Pomeron intercept $\alpha_P (0)=1+\Delta=1.13$.

 We consider that all intermediate states have the same structure and therefore we use the Schwimmer scheme to include higher-order rescatterings \cite{Armesto:2006ph}. The suppression from shadowing in $AB$ collisions for particles produced at rapidity $y$
 is obtained replacing the nuclear profile function $T_{AB}(b)$ by the integral on $\mathbf{s}$ of
\begin{equation}
S^{sh}(\mathbf{s},\mathbf{b}) = 
{T_A\left(\mathbf{s}+\frac{\mathbf{b}}{2}\right) \over 1 + A \ F(y,p_T)T_A\left(\mathbf{s}+\frac{\mathbf{b}}{2}\right)} \ {T_B\left(\mathbf{s}-\frac{\mathbf{b}}{2}\right) \over 1 + B\ F(-y,p_T) T_B\left(\mathbf{s}-\frac{\mathbf{b}}{2}\right)} 
\label{shad}
\end{equation}
where
\begin{equation}
F(y,p_T) = C \left [ \exp (\Delta \times y_{max}) - \exp (\Delta \times y_{min}\right ] /\Delta
\end{equation}
is the triple Pomeron graph contribution, with the rapidity of the triple Pomeron vertex integrated up to $y$ (where the trigger particle is produced), {\it i.e.} up to $y_{max} = y +  \ln{(\frac{\sqrt{s}}{m_T})}$. $y$ is the center of mass rapidity of the produced particle, $y>0$ for the projectile hemisphere (forward rapidity) and $y<0$ for 
the target one (backward rapidity). This integration limit is related to the parton model for hard processes: for projectile $A$ (target $B$),
$x_{A(B)}=\frac{m_T}{\sqrt{s}} e^{\pm y}$. $m_T$ is the transverse mass of the particle, $m_T=\sqrt{m^2+p_T^2}$. We take $y_{min} = \ln{(\frac{R_A m_N}{\sqrt{3}})} \sim \ln{(\frac{s}{M^{2}_{max}})} $. 
Here $m_N$ is the nucleon mass and $R_A = 0.82~A^{1/3} + 0.58$~fm is the Gaussian nuclear radius. $T_A$ and $T_B$ are the nuclear profile functions for which, as mentioned above, we use a Woods-Saxon parametrization. Note that the amount of shadowing depends on the nature and the transverse momentum of the produced particle through its transverse mass.

According to eq.~(\ref{shad}), the shadowing corrections
modify the A–dependence of the Glauber approximation for inclusive spectra in such a way that the behaviour 
$d \sigma_{AB}/dy \sim AB$ of the Glauber approximation changes to $d \sigma_{AB}/dy \sim A^{\alpha} B^{\alpha}$ where $\alpha < 1$. Let us remember the average number of collisions integrated over impact parameter typical for the Glauber model, $n_{coll}^{AB}=\frac{AB\sigma^{in}_{NN}}{\sigma^{in}_{AB}}$. For collisions of identical nuclei this leads to the well-known $A^{4/3}$ dependence.
The inclusive spectra of hadrons produced in the central rapidity region in nucleus-nucleus interactions can be obtained from the Glauber model
 applying the AGK cancellation. The multiplicity is in this case proportional to the number of cut Pomerons, leading
to a similar behaviour, $\frac{dN^{AA}}{dy}=\frac{A^2\sigma^{in}_{NN}}{\sigma^{in}_{AA}}\frac{dN^{pp}}{dy}=n_{coll}^{AA}\frac{dN^{pp}}{dy}$ for the yields of produced particles in the central rapidity region at high energies. 
Comparison with data shows that this relation significantly overestimates the hadron multiplicity at LHC energies.
 The corrections due to shadowing effects, arising by the contribution of the triple Pomeron graphs, cure this discrepancy and change the dependence of the multiplicity which behaves that way proportional to $A^{\delta}$. The value of delta is a  function of energy and it is equal to $\delta \simeq 1.1$ at LHC energies. 
Within this approach, a good description of the multiplicities from the Relativistic Heavy-Ion Collider (RHIC) to LHC energies is obtained \cite{Capella:2011vi}.

\section{Energy density and temperature of the medium}
\label{sec:temp}
In order to get the energy density and the temperature of the medium, we will carry out the following approach.
The energy deposition in the transverse plane corresponds to the transverse mass of the produced particles, so it is proportional to the multiplicities.
These are known to have two contributions, one proportional to the number of participant nucleons $N_{part}$ and the other, dominant at high energies, which is proportional to the number of binary collisions $N_{coll}$. According to the Glauber model, the number of collisions scales as $N_{part}^{4/3}$ for collisions of two equal nuclei.
On the other hand, the shadowing effects need to be taken into account.

As a first approach, in order to get the initial energy density one could take it as proportional to the density of wounded nucleons $N_{part}$. 
While at low energies this is a common assumption, the second contribution is expected at RHIC and LHC energies, proportional to the number of collisions.
However, at RHIC energies, the PHOBOS collaboration has checked at different energies (20, 60 and 200 GeV) that the total charged particle multiplicity exhibits
an essentially linear dependence on $N_{part}$ \cite{Alver:2010ck}.
Comparison of the averaged and scaled 200 GeV AuAu data with LHC data on multiplicities versus centrality 
shows a remarkable similarity in the shape of both distributions, at least up to peripheral collisions.
The fact that the shape of the normalised multiplicity distribution varies little with energy and stays almost constant up to TeV energies --even if the amount of hard processes, which scale with the number of binary collisions $N_{coll}$, is expected to contribute significantly to particle production at LHC and should lead to a steeper centrality dependence-- can be explained by a strong impact-parameter dependent shadowing of the nPDFs that limits this rise with centrality. 

\begin{figure}
\includegraphics{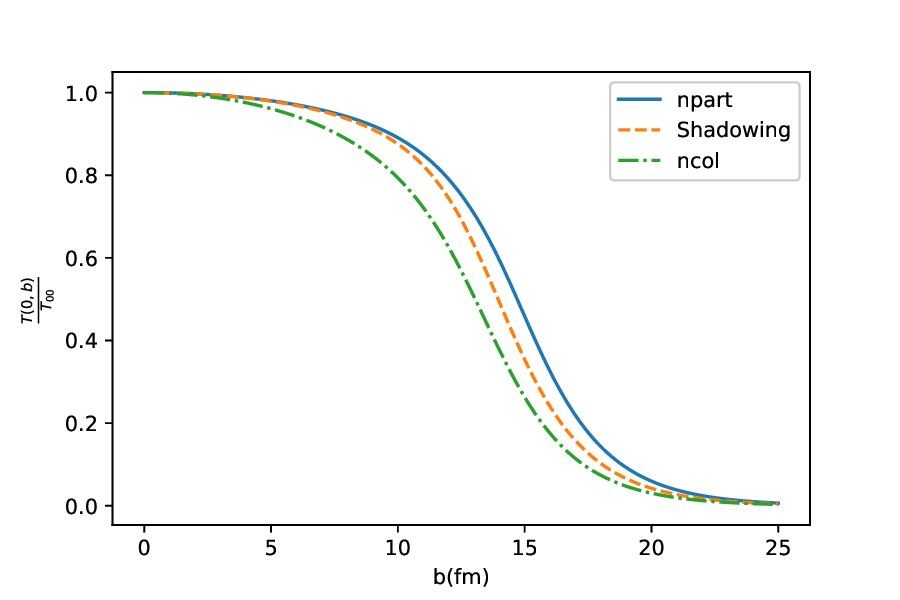}
\caption{Comparison of the temperature calculated in three different approaches -see main text for details.}
\label{fig:Tcompare}
\end{figure}

This is illustrated in Fig.~\ref{fig:Tcompare}, where the initial temperature at the center of the colliding region, taken as proportional to the energy density, $T_0(\mathbf{s},\mathbf{b}) \propto (\epsilon_i(\mathbf{s},\mathbf{b}))^{\frac{1}{4}}$, is shown. Three different approaches for the energy density have been used: proportional to the number of participants, proportional to the number of collisions and proportional to the multiplicity calculated based on the number of collisions corrected by the shadowing effects. Our results show that, up to peripheral collisions, the simplification that consists on taking the energy density proportional to the number of participants is correct and offers the same result as the one based on the multiplicity. In the following, we will use the results based on the multiplicities, which includes shadowing, to determine the initial temperature. In practice, the numerical cost of determining the temperature considering pion shadowing is almost equal to that of assuming that the temperature goes like the local number of participants to $1/4$.
 
\section{Application to quarkonium suppression}
\label{sec:ahq}
In this section, we discuss how the previous results could be used to compute the nuclear modification factor $R_{AA}$ of quarkonium in a heavy-ion collision. $R_{AA}$ is the quantity of a particular quarkonium state observed in a heavy-ion collision divided by the number that we would naively expect from proton-proton data. More precisely, in absence of any cold or hot nuclear matter effect any collision of nucleons would be approximately equivalent to a proton-proton collision. Then, since the number of collisions is proportional to $T_{AB}(b)$, $R_{AB}$ can be computed with
\begin{equation}
R_{AB}(b)=\frac{N^{AB}_{HQ}(b)}{N^{pp}_{HQ}T_{AB}(b)}\,,
\label{eq:raa}
\end{equation}
where $N^{AB}_{HQ}$ is the number of quarkonium states observed in heavy-ion collisions -for simplicity $N_{HQ}$- and $N^{pp}_{HQ}$ is the cross section of quarkonium in $pp$ collisions.

We can estimate $N_{HQ}$
as
\begin{equation}
N_{HQ}(b)=\int\,d^2sS^{sh}_{HQ}(\mathbf{s},\mathbf{b})S_{med}(\mathbf{s},\mathbf{b})\,
\label{eq:nhq}
\end{equation}
where $S^{sh}_{HQ}$ corresponds to the density of binary collisions corrected by the amount of shadowing on quarkonium estimated according to the model discussed in section \ref{sec:model} and $S_{med}$ is the medium induced suppression. In this way, if we assume that $S_{med}=1$, we would be taking into account only initial nuclear matter effects. On the other hand, we can ignore initial nuclear matter effects by substituting $S^{sh}_{HQ}(\mathbf{s},\mathbf{b})$ by $T_{AB}(\mathbf{s},\mathbf{b})$, {\it i.e.} taking $F(y)$ in eq.~(\ref{shad}) equal to 0. We name the quantity obtained by setting $S_{med}=1$, $R_{AA}^{I}$. Instead, if we substitute $S^{sh}_{HQ}(\mathbf{s},\mathbf{b})$ by $T_{AB}(\mathbf{s},\mathbf{b})$ we obtain $R_{AA}^T$. Note that we are assuming that quarkonium is comoving with the medium. To consider finite velocity effects is out of the scope of this work and is left for future investigations. A quarkonium that is not comoving with the medium would be sensitive to the temperature of different medium regions at different times. The initial condition model we are using would still be valid in this situation, however, eq. (\ref{eq:nhq}) would not be valid anymore.

The effects of the hot medium are usually taken as $S_{med}(\mathbf{s},\mathbf{b})=f\left(T_0(\mathbf{s},\mathbf{b})\right)$, where $T_0$ is the initial temperature of the medium. In other words, the medium suppression depends on the initial temperature at the point in which the quarkonium pair is created. 
Taking the initial energy density proportional to the number of pions created at a given point
the calculation of the initial temperature is straightforward. A common assumption is
\begin{equation}
T_0(\mathbf{s},\mathbf{b}) \propto (\epsilon_i(\mathbf{s},\mathbf{b}))^{\frac{1}{4}}
\end{equation}
where we have considered that the equation of state is that of a free gas. We assume that the energy density at a given point is proportional to the production of pions. Then, the temperature just after the collision is given by 
\begin{equation}
T_0(\mathbf{s},\mathbf{b})=T_{00}\left(\frac{S^{sh}_\pi(\mathbf{s},\mathbf{b})}{S^{sh}_\pi(\mathbf{0},\mathbf{0})}\right)^{1/4}\,,
\label{eq:T0}
\end{equation}
where, by construction, $T_{00}$ is the temperature in the center of the overlapping region for the most central collisions \footnote{Note that both $S^{sh}_{HQ}$ and $S^{sh}_\pi$ correspond to eq.~(\ref{shad}). They differ in the value of $F(0)$, since in the first case we use the value of $m_T$ that corresponds to bottomonium and in the former case we use a value suitable for pions. }. This is a quantity whose computation is out of the scope of this work.
It can be adjusted using experimental results within a hydrodynamical framework \cite{Alberico:2013bza,Alqahtani:2020paa}. The value for PbPb collisions at $\sqrt{s}=5.02$ TeV is estimated to be 502 MeV.
We are going to focus on computing the relation of the temperature at different points and values of the impact parameter with $T_{00}$. We can compute the average initial temperature seen by a quarkonium state in a given collision, $\langle T_{HQ}\rangle$, by
\begin{equation}
\langle T_{HQ}(b)\rangle=T_{00}\frac{\int\,d^2sS^{sh}_{HQ}(\mathbf{s},\mathbf{b})\left(\frac{S^{sh}_\pi(\mathbf{s},\mathbf{b})}{S^{sh}_\pi(\mathbf{0},\mathbf{0})}\right)^{1/4}}{\int\,d^2sS^{sh}_{HQ}(\mathbf{s},\mathbf{b})}\,.
\end{equation}
\begin{figure}
\includegraphics{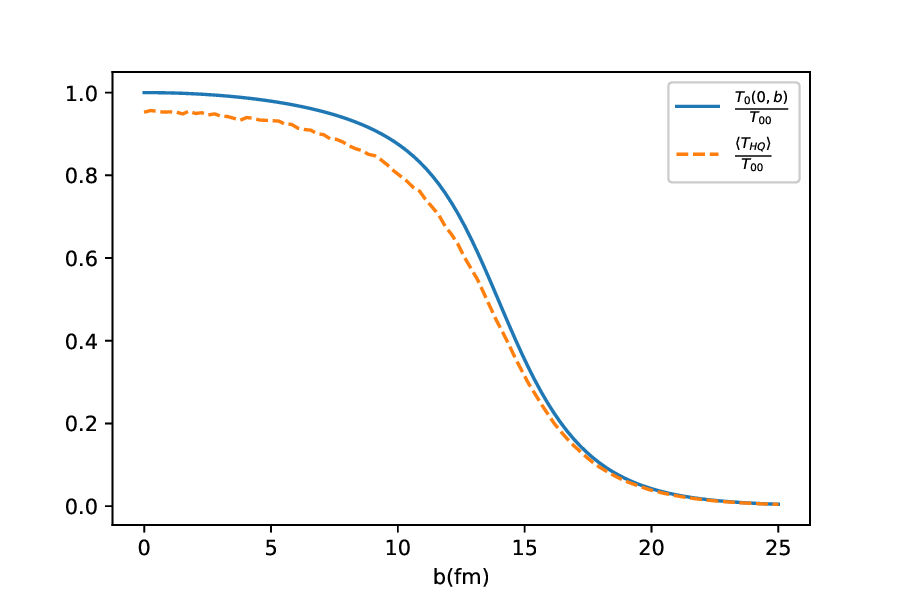}
\caption{Comparison between $T_0(\mathbf{0},\mathbf{b})$ and $\langle T_{HQ}(b)\rangle$}
\label{fig:comT}
\end{figure}
A good approximation to the initial temperature in a not too peripheral collision is given by $T_0(\mathbf{0},\mathbf{b})$. The reason is that $T_{AB}$ has a central plateau in which it is approximately constant. We compare $T_0(\mathbf{0},\mathbf{b})$ and $\langle T_{HQ}(b)\rangle$ in Fig.~\ref{fig:comT}. We see indeed that to take the temperature at the center of the overlapping region as representative of the temperature of all the collision is a reasonable approximation, at least regarding the temperature seen by a quarkonium state.

In Table \ref{tab:hq} we show the numerical values of some quantities discussed in our review of the Glauber model of interest for the study of quarkonium suppression. Among them, we computed $R_{AA}^{I}$ and $\langle T_{HQ}\rangle$, both quantities discussed in this section. All the computations were done for the conditions met in LHC PbPb collisions at $\sqrt{s}=5.02\,\rm{TeV}$ for upsilon production and using the model discussed in section \ref{sec:model}.
\begin{table}
\begin{tabular}{|c|c|c|c|c|}
\hline
$b$ (fm) & $N_{part}$ & $N_{coll}$ & $R_{AA}^{I}$ & $\frac{\langle T_{HQ}\rangle}{T_{00}}$\\
\hline
$0$ & $405\pm 1$ & $(211\pm 1)\cdot 10$ & $0.649\pm 0.003$ & $0.954\pm 0.002$ \\
$1$ & $401\pm 1$ & $(206\pm 1)\cdot 10$ & $0.645\pm 0.003$ & $0.958\pm 0.002$ \\
$2$ & $387\pm 1$ & $(194\pm 1)\cdot 10$ & $0.651\pm 0.003$ & $0.950\pm 0.002$ \\
$3$ & $361\pm 1$ & $(173\pm 1)\cdot 10$ & $0.651\pm 0.004$ & $0.947\pm 0.002$ \\
$4$ & $328\pm 1$ & $(151\pm 1)\cdot 10$ & $0.658\pm 0.004$ & $0.943\pm 0.001$ \\
$5$ & $289\pm 1$ & $(126\pm 1)\cdot 10$ & $0.665\pm 0.004$ & $0.930\pm 0.002$ \\
$6$ & $249\pm 1$ & $(101\pm 1)\cdot 10$ & $0.670\pm 0.005$ & $0.920\pm 0.002$ \\
$7$ & $207\pm 1$ & $767\pm 4$ & $0.676\pm 0.006$ & $0.901\pm 0.003$ \\
$8$ & $166\pm 1$ & $558\pm 3$ & $0.704\pm 0.007$ & $0.870\pm 0.003$ \\
$9$ & $127\pm 1$ & $378\pm 2$ & $0.720\pm 0.008$ & $0.845\pm 0.003$ \\
$10$ & $91\pm 1$ & $231\pm 2$ & $0.74\pm 0.01$ & $0.809\pm 0.003$ \\
\hline
\end{tabular}
\caption{Values of several quantities relevant for the study of quarkonium suppression in heavy-ion collisions of the type performed at LHC ($\sqrt{s}=5.02\,\rm{TeV}$). The quantities are defined in the text and are computed using Monte Carlo integration.  The errors correspond to the statistical error of the integration.}
\label{tab:hq}
\end{table}

\section{Illustration of how to include shadowing in a $R_{AA}$ computation}
\label{sec:raa}
In what follows, we will show the applicability of our approach within several models of the quarkonium-medium interaction. 
In the previous sections we have discussed how $R_{AA}$ can be computed ignoring medium effects by taking eq.~(\ref{eq:nhq}) in the case $S_{med}=1$. The computation of the survival probability of quarkonium in medium is
a very active research topic in itself. In this section we discuss the results of two different scenarios. In the first one, we make use of a simplified model inspired by recent potential non-relativistic QCD computations \cite{Brambilla:2016wgg,Brambilla:2020qwo}. In the second one, we use the gap model developed in \cite{Blaizot:2021xqa} that implements the lattice static potential.
\subsection{pNRQCD inspired model}
\label{ssec:pNRQCD}
Let us introduce a simple model for the computation $S_{med}(\mathbf{s},\mathbf{b})$. Our purpose here is not to produce an accurate and rigorous prediction of $R_{AA}$, we just aim to illustrate a procedure in which initial and hot nuclear matter effects can be included assuming that $S_{med}$ is known.

If the corrections to the binding energy and the decay width are much smaller than the binding energy in the vacuum, the evolution of quarkonium can be modeled with a rate equation \cite{Blaizot:2018oev}. Moreover, in the large $N_c$ limit and as long as the population of octets is not much bigger than that of singlets, we can ignore recombination effects from octets decaying into singlets \cite{Escobedo:2020tuc}. Then, the survival probability of a quarkonium state is given by
\begin{equation}
S_{med}(\mathbf{s},\mathbf{b})=e^{-\int_{t_0}^{t_f(\mathbf{s},\mathbf{b})}\,d\tau\Gamma(\mathbf{s},\mathbf{b},\tau)}\,.
\end{equation}
Let us also assume that the temperature is such that the medium sees quarkonium as a small dipole \cite{Brambilla:2016wgg}. In this limit $\Gamma(T)=\kappa \langle r^2\rangle$, where $\kappa$ is proportional to a gauge invariant expectation value involving only gauge fields. If QCD was a conformal theory, then $\kappa$ would scale like $T^3$. For simplicity, we are going to assume that $\kappa$ scales like that, although this assumption is not justified by lattice data \cite{Brambilla:2020siz}. In fact, considering a non-trivial dependence of $\Gamma$ with the temperature in the small dipole limit seems to improve agreement with data \cite{Brambilla:2020qwo}.

Regarding the evolution of the temperature with time, we are going to assume Bjorken evolution approximating the speed of sound by that of a free gas. Under this assumption $tT^3$ is a constant, and so is $t\Gamma(t)$. Therefore,
\begin{equation}
\int_{t_0}^{t_f(\mathbf{s},\mathbf{b})}\,d\tau\Gamma(\mathbf{s},\mathbf{b},\tau)=t_0\Gamma(\mathbf{s},\mathbf{b},t_0)\int_{t_0}^{t_f(\mathbf{s},\mathbf{b})}\frac{\,d\tau}{\tau}=t_0\Gamma(\mathbf{s},\mathbf{b},t_0)\log\left(\frac{t_f(\mathbf{s},\mathbf{b})}{t_0}\right)\,.
\end{equation}
We assume that $t_0$ is always the same for a given collision type and we define $t_f$ as the time at which quarkonium arrives to a given temperature $T_f$ below which we no longer consider thermal effects. For a medium following Bjorken evolution
\begin{equation}
\left(\frac{t_f(\mathbf{s},\mathbf{b})}{t_0}\right)=\left(\frac{T_0(\mathbf{s},\mathbf{b})}{T_f}\right)^3\,.
\end{equation}
Then,
\begin{equation}
S_{med}(\mathbf{s},\mathbf{b})=\begin{cases}
\left(\frac{T_f}{T_0(\mathbf{s},\mathbf{b})}\right)^{3t_0\Gamma(\mathbf{s},\mathbf{b},t_0)} & T_0(\mathbf{s},\mathbf{b})\geq T_f\\
1 & T_0(\mathbf{s},\mathbf{b})< T_f
\end{cases}
\end{equation}
Let us define $\alpha=\frac{T_f}{T_{00}}$ and $\beta=\frac{3t_0\Gamma(\mathbf{s},\mathbf{b},t_0)T_{00}^3}{T_0^3(\mathbf{s},\mathbf{b})}$. Note that, because $\Gamma$ scales like $T^3$, $\beta$ is a constant that does not depend on $b$ and $s$. Then we can rewrite,
\begin{equation}
S_{med}(\mathbf{s},\mathbf{b})=\begin{cases}
\left(\frac{\alpha T_{00}}{T_0(\mathbf{s},\mathbf{b})}\right)^{\beta \left(\frac{T_0(\mathbf{s},\mathbf{b})}{T_{00}}\right)^3} & T_0(\mathbf{s},\mathbf{b})\geq \alpha T_{00} \\
1 & T_0(\mathbf{s},\mathbf{b})<\alpha T_{00}
\end{cases}
\label{eq:smedc}
\end{equation}
This is the formula that we are going to use in our computations of thermal effects in this subsection. $\alpha$ and $\beta$ have been chosen such to qualitatively follow experimental results. Let us remark again that our purpose here is not to make a state-of-the-art description of quarkonium in a medium but rather to have a simple analytic formula for quarkonium's survival probability that we can use to illustrate how to include cold nuclear matter effects in a thermal model within our approach.
\begin{figure}
\includegraphics[scale=1]{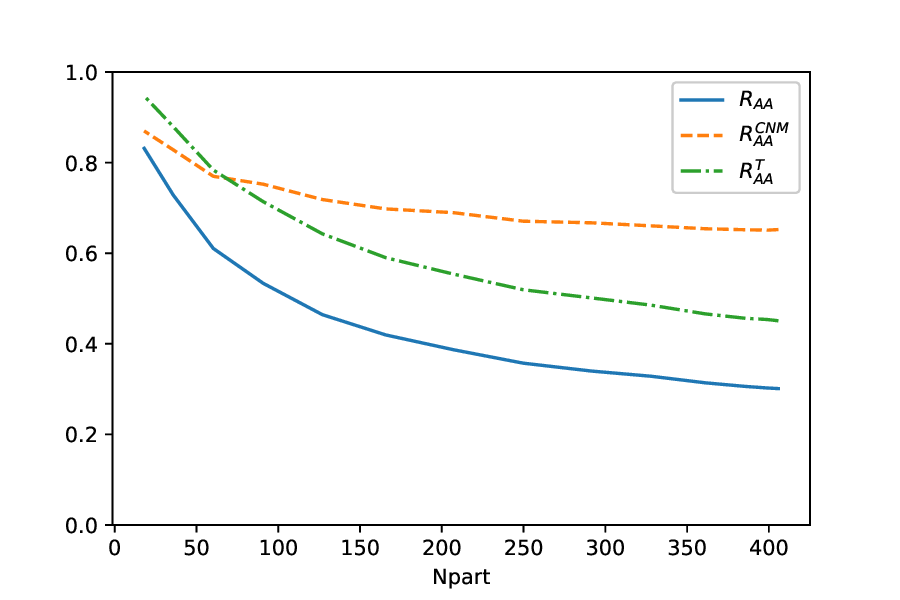}
\caption{Computation of $R_{AA}$ using the model discussed in subsection \ref{ssec:pNRQCD}. $R_{AA}^{CNM}$ represents the initial state effects while $R_{AA}^T$ represent the medium effects.}
\label{fig:raa}
\end{figure}

In order to compute $R_{AA}$ we use eqs.~(\ref{eq:raa}), (\ref{eq:nhq}) and (\ref{eq:T0}). In Fig.~\ref{fig:raa}, we show our results for $R_{AA}$, $R_{AA}^{CNM}$ (which is considered to include our initial-state effects) and $R_{AA}^T$ (see definitions in section \ref{sec:ahq}) using eq.~(\ref{eq:smedc}) with $\alpha=0.4$ and $\beta=1.2$. Although the model is very naive, the computation allows us to obtain some conclusions. Shadowing effects are significant, although smaller than medium effects. However, they have some significant qualitative differences. Shadowing effects have a milder dependence with the number of participants and they decrease slowly as this number goes to zero. It is also important to note that the total suppression is not the sum of the suppression due to shadowing plus the suppression due to in medium effects. This implies that the influence of shadowing on $R_{AA}$ depends on the specific model for quarkonium-medium interaction, at least if the aim is to go beyond qualitative considerations. 

We remark again that the survival probability used in this subsection is a simplified model based on potential NRQCD (pNRQCD) arguments valid in the region $\frac{1}{r}\gg T\gg E$. The parameters in eq.~(\ref{eq:smedc}) were chosen in an {\it ad hoc} way to qualitatively reproduce experimental results on $R_{AA}$ once shadowing effects are taken into account. A more precise, rigorous and state-of-the-art application of the pNRQCD framework to the computation of $R_{AA}$, ignoring shadowing effects, can be found in \cite{Brambilla:2020qwo,Brambilla:2021wkt}.

\subsection{Gap model}
\label{ssec:jp}
In this subsection we aim to include shadowing effects in the model developed in \cite{Blaizot:2021xqa}. More specifically, we focus on the scenario discussed in \cite{Blaizot:2021xqa} in which lattice QCD data on the static potential is used to compute the decay width of $\Upsilon(1S)$. The emphasis of \cite{Blaizot:2021xqa} was to highlight the importance of taking into account the finite energy gap between color singlet and color octets states. The decay width of quarkonium is suppressed compared to computations which do not take into account the finite value of the energy gap. This is a remarkable effect, especially at small temperatures. In \cite{Blaizot:2021xqa}, it was shown that suppression is overestimated by an approximate factor of two when the gap is ignored. 

The $S_{med}$ that we are going to use in this section is
\begin{equation}
    S_{med}(\mathbf{s},\mathbf{b})=\begin{cases}\mathrm{e}^{-\frac{3aT_0(\mathbf{s},\mathbf{b})^3t_0}{b^2}\left(\mathrm{e}^{-\frac{b}{T_0(\mathbf{s},\mathbf{b})}}\left(1+\frac{b}{T_0(\mathbf{s},\mathbf{b})}\right)-\mathrm{e}^{-\frac{b}{T_f}}\left(1+\frac{b}{T_f}\right)\right)} & T_0(\mathbf{s},\mathbf{b})> T_f
    \\
    1 & T_0(\mathbf{s},\mathbf{b})\leq T_f
\end{cases}
    \label{eq:spjp}
\end{equation}
where $a=22.9$, $b=2170\,\rm{MeV}$, $t_0=0.6\,\rm{fm}$, $T_f=175\,\rm{MeV}$ and $T_0$ is the initial temperature. Details about the derivation of this formula can be found in \cite{Blaizot:2021xqa}. Here we just mention briefly some features of the model. The model consists of a rate equation, derived from an open quantum system approach in the limit $E\gg \Gamma$. The decay width takes into account the finite energy gap between singlets and octets. Lattice QCD data on the static potential is used to determine the wave function and the binding energies of quarkonium, both information is needed to compute the decay width. In \cite{Blaizot:2021xqa}, the decay width was computed for several temperatures. The results could be well fitted by the following function
\begin{equation}
    \Gamma(T)=aT\mathrm{e}^{-\frac{b}{T}}\,.
\end{equation}
This formula was then used to derive eq. (\ref{eq:spjp}). Here we improve the previous computation by including shadowing effects. They influence $R_{AA}$ in two different ways. By modifying the value of $T_0$ in eq.~(\ref{eq:spjp}) and by the shadowing effects included in $S^{sh}_{HQ}$ at eq.~(\ref{eq:nhq}). 
\begin{figure}
\includegraphics[scale=1]{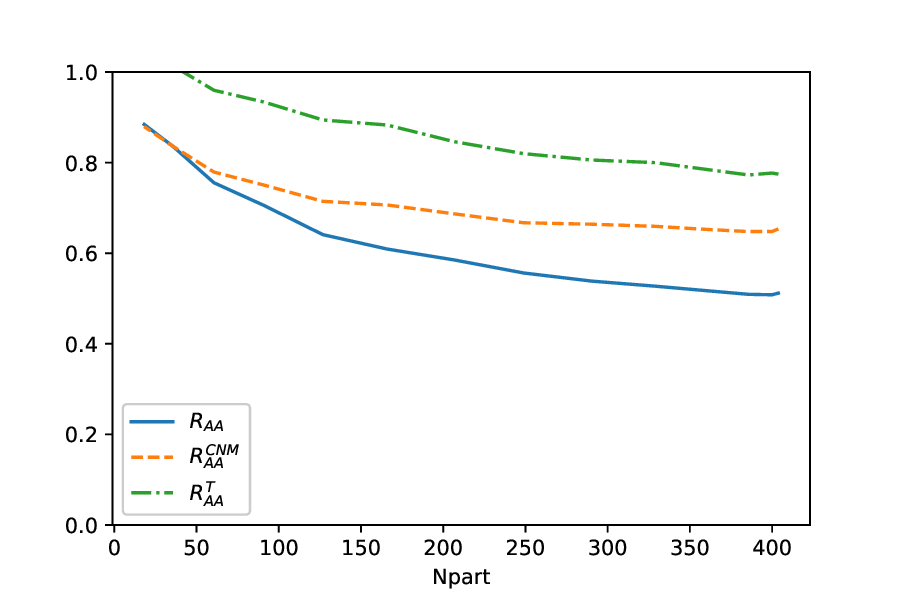}
\caption{Computation of $R_{AA}$ using the model discussed in subsection \ref{ssec:jp}. $R_{AA}^{CNM}$ represents the initial state effects while $R_{AA}^T$ represent the medium effects.}
\label{fig:raa2}
\end{figure}

Our results are shown in Fig.~\ref{fig:raa2}. We note that $R_{AA}^T$ is almost identical to the $R_{AA}$ results obtained in \cite{Blaizot:2021xqa}. This is the expected result since the only difference is that in the computation of $R_{AA}^T$ pion shadowing was taken into account while in the \cite{Blaizot:2021xqa} it was assumed that the energy density scales like the local density of participants. This is, as we discussed, a very good approximation at LHC energies. Note also that $R_{AA}^{CNM}$ comes from the same formulas in Figs.~\ref{fig:raa} and \ref{fig:raa2}, the only difference is due to statistical fluctuations in the Monte Carlo integration done to obtain the results. Focusing on $R_{AA}$, including both shadowing and medium effects, we observe that shadowing effects are practically the only source of suppression when the number of participants is below $50$. Our interpretation is that at very small temperatures the energy gap suppresses the decay width very strongly, leaving shadowing as the only source of suppression. Finally, the results in Fig.~\ref{fig:raa2} reinforce the observation that the overall suppression is not simply the sum of the suppression due to shadowing plus the suppression due to thermal medium effects.

\section{Conclusions}
\label{sec:concl}
In this manuscript we have discussed how shadowing corrections can be easily included in the computation of quarkonium suppression. Our main assumptions are that the medium induced suppression can be described by a survival probability, the validity of the shadowing model discussed in \cite{Capella:2011vi}, that the initial temperature scales as $1/4$ of the initial energy density and that this quantity is dominated by the pions' contribution. We note that our method could in principle be used in combination with a Markovian quantum evolution of quarkonium in a medium (as that in the recent paper \cite{Brambilla:2021wkt}) as long as recombination effects are ignored.

One of our main focuses is the temperature felt by a quarkonium state formed at a given point in the medium. We found that to assume that the energy density is proportional to the density of participants gives almost the same result as considering a more realistic model of pion shadowing, at least regarding the initial temperature. Another interesting observation is that the average temperature seen by a quarkonium state is approximately equal to the temperature in the center of the overlapping region. 

Finally, we have applied our methods to two different models of the medium-quarkonium interaction. The first is a model inspired by recent pNRQCD results in which the medium sees quarkonium as a small color dipole and the temperature is much larger than the binding energy. The second is one of the scenarios considered in \cite{Blaizot:2021xqa}. In both cases we found that shadowing effects are smaller than medium effects but of the same order of magnitude. Shadowing has a milder dependence on the number of participants than medium induced suppression. Finally, it is interesting to observe that the overall suppression is not simply the sum of the suppression due to shadowing plus that due to medium effects.

In principle, our method could also be used with models in which medium effects are a mixture of quark-gluon plasma and hadron gas effects, as long as these effects can be encoded in a survival probability.

\section*{Acknowledgements}
We thank Nestor Armesto, Jean-Paul Blaizot and Nora Brambilla for useful discussions. We have received financial support from Xunta de Galicia, by European Union ERDF, by the "Mar\'{\i}a de Maeztu" Units of Excellence program MDM2016-0692, by the Spanish Research State Agency under projects PID2020-119632GB-I00 and FPA2017-83814-P and from
the European Research Council project ERC-2018-ADG-835105 YoctoLHC. We acknowledge the lively discussions that took place at the workshop "Suppression and (re)generation of quarkonium in heavy-ion collisions at the LHC" supported by the EMMI Rapid Reaction
Task Force developed in GSI Helmholtzzentrum für Schwerionenforschung in December 2019.
\bibliography{optglau}
\end{document}